\documentclass{article}
\usepackage{spconf,amsmath,graphicx}
\usepackage[fontsize=9pt]{scrextend}

\usepackage{booktabs}
\usepackage{balance}

\usepackage{amsmath,epsfig,amssymb,amsthm,url,amsfonts}
\usepackage{algorithm}% http://ctan.org/pkg/algorithm
\usepackage{algpseudocode}% http://ctan.org/pkg/algorithmicx
\usepackage{multirow}
\usepackage{mdframed}
\usepackage{url}
\usepackage{enumitem}
\usepackage[makeroom]{cancel}
\usepackage{dblfloatfix}
\usepackage{array}
\usepackage{comment}
\usepackage{epstopdf}
\usepackage{float}
\usepackage{subfig}
\usepackage{breqn}
\usepackage{cite}
\usepackage{xcolor}
\usepackage{cases}
\usepackage{tabularx}
\usepackage[printonlyused,withpage]{acronym}
\usepackage{balance}
\usepackage{graphbox} %loads graphicx package

\usepackage{hyperref}
\setlist[itemize]{label=$\triangleright$}
\newtheoremstyle{break}% name
{}%         Space above, empty = `usual value'
{}%         Space below
{\itshape}% Body font
{}%         Indent amount (empty = no indent, \parindent = para indent)
{\bfseries}% Thm head font
{.}%        Punctuation after thm head
{\newline}% Space after thm head: \newline = linebreak
{}%         Thm head spec
\theoremstyle{break}

\theoremstyle{definition}

\newcommand{\vect}[1]{\mathbf{#1}}
\newcommand{\bs}[1]{\boldsymbol{#1}}
\newcommand{\E}{\mathbb{E}}
\usepackage{url}

\newcommand{\argmax}{\operatornamewithlimits{argmax}}

\newcommand{\lk}{ \left\{ }
\newcommand{\rk}{ \right\} }

\newcommand{\Rbb}{{\mathbb{R}}}

\newcommand{\Hb}{{\bf H}}

\newcommand{\bb}{{\bf b}}

\newcommand{\Wb}{{\bf W}}

\newcommand{\diag}{\mbox{{diag}}}

\newsavebox\mybox

\acrodef{SE}{speech enhancement}
\acrodef{STFT}{short-time Fourier transform}
\acrodef{STOI}{short-time objective intelligibility}
\acrodef{PSD}{power spectral density}
\acrodef{NMF}{nonnegative matrix factorization}
\acrodef{AV}{audio-visual}
\acrodef{DNN}{deep neural network}
\acrodefplural{DNNs}{deep neural networks}
\acrodef{VAE}{variational auto-encoder}
\acrodefplural{VAEs}{variational auto-encoders}
\acrodef{CVAE}{conditional variational auto-encoder}
\acrodefplural{CVAEs}{conditional variational auto-encoders}
\acrodef{A-VAE}{audio VAE}
\acrodef{V-VAE}{visual VAE}
\acrodef{AV-CVAE}{audio-visual CVAE}
\acrodef{ROI}{region of interest}
\acrodef{MCMC}{Markov Chain Monte Carlo}
\acrodef{EM}{expectation-maximization}
\acrodef{MCEM}{Monte Carlo expectation-maximization}
\acrodef{TF}{time frequency}
\acrodef{ELBO}{evidence lower bound}
\acrodef{ROI}{region of interest}
\acrodef{LR}{Living Room}
\acrodef{SDR}{signal-to-distortion ratio}
\acrodef{PESQ}{perceptual evaluation of speech quality}
\acrodef{ASE}{audio speech enhancement}
\acrodef{VSE}{visual speech enhancement}
\acrodef{AVSE}{audio-visual speech enhancement}
\acrodef{SNR}{signal-to-noise ratio}
\acrodefplural{SNRs}{signal-to-noise ratios}
\acrodef{LSTM}{long short-term memory}
\acrodef{DNNs}{deep neural networks}

\title{A weighted-variance variational autoencoder model for speech enhancement}
\name{%
Ali Golmakani$^1$, %
Mostafa Sadeghi$^1$, %
Xavier Alameda-Pineda$^2$, and
Romain Serizel$^1$ \thanks{Experiments presented in this paper were carried out using the Grid'5000 testbed, supported by a scientific interest group hosted by Inria and including CNRS, RENATER, and several Universities as well as other organizations (see https://www.grid5000.fr).}
}
\address{%
$^1$Université de Lorraine, CNRS, Inria, LORIA, F-54000 Nancy, France\\ 
$^2$Inria Grenoble \& Univ. Grenoble Alpes, France}

\begin{document}

\maketitle
 
\begin{abstract}
We address speech enhancement based on variational autoencoders, which involves learning a speech prior distribution in the time-frequency (TF) domain. A zero-mean complex-valued Gaussian distribution is usually assumed for the generative model, where the speech information is encoded in the variance as a function of a latent variable. In contrast to this commonly used approach, we propose a \textit{weighted} variance generative model, where the contribution of each spectrogram time-frame in parameter learning is weighted. We impose a Gamma prior distribution on the weights, which would effectively lead to a Student's t-distribution instead of Gaussian for speech generative modeling. We develop efficient training and speech enhancement algorithms based on the proposed generative model. Our experimental results on spectrogram auto-encoding and speech enhancement demonstrate the effectiveness and robustness of the proposed approach compared to the standard unweighted variance model.
\end{abstract}

\section{Introduction}
\label{sec:intro}
Speech enhancement is a fundamental task in signal processing and machine learning, aiming to recover a clean speech signal from a noisy observation \cite{vincent2018audio}. A classical approach to this problem involves statistical modeling of clean speech and noise signals, e.g., using non-negative matrix factorization (NMF), followed by an inference method such as maximum likelihood (ML) or Maximum a posteriori (MAP) estimation \cite{fevotte2009nonnegative,smaragdis2007supervised}. However, with the advent of deep learning, there has been a significant shift towards supervised (discriminative) frameworks, which train a \ac{DNN} on a large collection of paired clean and noisy speech signals \cite{wang2018supervised}. Nevertheless, these methods suffer from generalization issues, e.g., for unseen noise environments, as the train and test conditions might be significantly different.

Recently, there has been a growing interest in alternative approaches based on deep generative models, including variational autoencoders (VAEs) \cite{bando2018statistical,leglaive2018variance,sadeghi2020audio,bie2022unsupervised,fang2021variational}, generative adversarial networks (GANs) \cite{pascual2017segan}, and normalizing flows (NFs) \cite{nugraha2020flow}, due to their \textit{potential} generalization advantage. In particular, VAE-based speech enhancement involves learning a prior distribution of (time-frequency domain) clean speech data with a latent variable model. More precisely, the distribution of each speech time-frequency (TF) point is modeled as a circularly symmetric complex Gaussian distribution, where the variance is a \ac{DNN}-based parameterized function of a latent variable with a standard Gaussian prior. Given a noisy speech observation and the trained speech prior, a parametric statistical noise model is adaptively learned with an \ac{EM} approach followed by clean speech signal estimation. Therefore, noise characteristics are modeled at test time, giving them higher potential for performance generalization compared to supervised methods \cite{bando2018statistical,leglaive2018variance}.

In this paper, we propose to use a \textit{weighted} variance circularly symmetric complex Gaussian distribution for VAE-based speech modeling, where the contribution of each spectrogram time-frame to parameter learning and inference is separately weighted. Assuming a Gamma prior for the weights and marginalizing them, the resulting model would become a Student's t-distribution. This brings more efficient, robust, and flexible modeling power than the standard unweighted Gaussian variance model. We develop computationally efficient training and speech enhancement methods based on the EM framework. Our experiments show that the proposed weighted variance VAE model outperforms the standard unweighted counterpart, both in terms of reconstruction quality and speech enhancement performance.

The rest of the paper is organized as follows. Section~\ref{sec:vae} reviews VAE-based speech generative modeling. The proposed speech generative and enhancement frameworks are detailed in Section~\ref{sec:prop}. Section~\ref{sec:related} discusses the related work. Experimental results are presented in Section~\ref{sec:exp}. Finally, Section~\ref{sec:conc} concludes the paper.

\section{VAE-based Speech modeling}
\label{sec:vae}

We denote the \ac{STFT} representation of clean speech signals as $\vect{s}=\{\vect{s}_1,\ldots,\vect{s}_T\}$, consisting of complex-valued vectors ${\vect{s}_t = [s_{ft}]_{f=1}^F \in \mathbb{C}^F}$. The VAE framework associates a latent variable $\vect{z}_t\in \mathbb{R}^L$ to each time frame $\vect{s}_t$, where ${L\ll F}$. The joint distribution, that is $p{(\vect{s}_t, \vect{z}_t) = p(\vect{s}_t| \vect{z}_t)\cdot p(\vect{z}_t)}$, is modeled by some parametric Gaussian forms:
\begin{equation}\label{eq:genmodel}
p_{\theta}(\vect{s}_t| \vect{z}_t) = \mathcal{N}_c\Big(\bs{0}, \diag(\bs{\sigma}_{\theta}^2(\vect{z}_t))\Big),~~~p(\vect{z}_t) = \mathcal{N}(\bs{0}, \vect{I}),
\end{equation}
where $\mathcal{N}_c(\bs{0}, \bs{\Sigma})$ denotes a circularly symmetric complex Gaussian distribution, and $\vect{I}$ is the identity matrix. Also,  $\bs{\sigma}_{\theta}(.)$ (applied element-wise) is a non-linear function denoting the standard deviation, which is modeled by some \ac{DNN}, called the \textit{decoder},  with parameters $\theta$. To learn $\theta$, one needs to compute the posterior distribution $p_{\theta}(\vect{z}_t| \vect{s}_t)$, which is intractable. In the VAE framework, this term is approximated as follows:
\begin{equation}\label{eq:encoder}
	q_{\psi}(\vect{z}_t| \vect{s}_t) = \mathcal{N}(\bs{\mu}_{\psi}(\vect{s}_t), \diag(\bs{\sigma}_{\psi}^2(\vect{s}_t)),
\end{equation}
where $\bs{\mu}_{\psi}$ and $\bs{\sigma}_{\psi}$ are implemented using a \ac{DNN}, called the \emph{encoder}.

The set of parameters, $\Phi=\{\theta, \psi\}$, is learned by optimizing a lower-bound, denoted $\mathcal{L}(\Phi;\vect{s})$, on the intractable data log-likelihood $\log p_{\theta}(\vect{s})$. This is achieved by defining the evidence lower-bound (ELBO) as follows \cite{KingW14}:
\begin{equation}\label{eq:elbo}
\mathcal{L}(\Phi;\vect{s}) = \E_{q_{\psi}(\vect{z}| \vect{s}) }\lk\log p_{\theta}(\vect{s}| \vect{z})\rk - \mathcal{D}_{\textsc{kl}}(q_\psi(\vect{z}|\vect{s}) \|p(\vect{z})), 
\end{equation}
where $\mathcal{D}_{\textsc{kl}}(q \|p)$ stands for the Kullback–Leibler (KL) divergence between $q$ and $p$. The first term in \eqref{eq:elbo} measures the reconstruction quality of the model, and the second one is a regularization. Training proceeds by optimizing $\mathcal{L}(\Phi;\vect{s}) $ over $\Phi$ using a gradient-based optimizer, along with the \emph{reparametrization trick} \cite{KingW14}.

\section{Proposed framework}
\label{sec:prop}
\subsection{Generative model}
\label{sec:model}
As opposed to the commonly used unweighted variance model presented in \eqref{eq:genmodel}, we propose a more flexible distribution, by introducing some weight parameters $w_t>0$:
\begin{equation}\label{eq:propgenmodel}
\begin{cases}
p_{\theta}(\vect{s}_t| \vect{z}_t, w_t) = \mathcal{N}_c\Big(\bs{0}, \diag(\bs{\sigma}_{\theta}^2(\vect{z}_t))/w_t\Big),\\
p(\vect{z}_t) = \mathcal{N}(\bs{0}, \vect{I}),\\
p(w_t) = \mathcal{G}(w_t;\alpha,\beta),
\end{cases}
\end{equation}
where, $\mathcal{G}(w;\alpha,\beta)$ is the Gamma distribution ($\alpha,\beta>0$):
\begin{equation}
    \mathcal{G}(w;\alpha,\beta) = \frac{\beta^\alpha}{\Gamma(\alpha)} w^{\alpha-1} \exp(-\beta w),
\end{equation}
and $\Gamma(.)$ denotes the gamma function. The mean and variance of this distribution are equal to $\alpha/\beta$ and $\alpha/\beta^2$, respectively. Note that $p_{\theta}(\vect{s}_t| \vect{z}_t)$ is essentially an infinite mixture of Gaussian distributions: $p_{\theta}(\vect{s}_t| \vect{z}_t)= \int p_{\theta}(\vect{s}_t| \vect{z}_t, w_t) p(w_t) \textrm{d} w_t$. This effectively takes the form of a Student's t-distribution, which is well-known for its robustness and flexibility advantages over a standard Gaussian distribution \cite{lange1989robust}.

\subsection{Parameters inference}
\label{sec:inference}
To learn the parameters of the proposed Student VAE (St-VAE) model, denoted $\widetilde{\Phi}=\{\theta, \psi, \alpha, \beta \}$, we need to compute the posterior distribution of the latent variables $\vect{z}_t, w_t$:

\begin{equation}
    p_{\theta}(\vect{z}_t, w_t | \vect{s}_t) = p_{\theta}(w_t | \vect{s}_t, \vect{z}_t)\cdot p_{\theta}(\vect{z}_t| \vect{s}_t).
\end{equation}
The first term writes $p_{\theta}(w_t | \vect{s}_t, \vect{z}_t)\propto p_{\theta}(\vect{s}_t|\vect{z}_t, w_t)\cdot p(w_t) = \mathcal{G}(\alpha'_t,\beta'_t)$, where:

\begin{equation}
    \begin{cases}
        \alpha'_t = \alpha + F\\
        \beta'_t = \beta + \sum_{f} \frac{|s_{ft}|^2}{{\sigma}_{\theta, f}^2(\vect{z}_t)}.
    \end{cases}
\end{equation}
The second posterior distribution, i.e., $p_{\theta}(\vect{z}_t| \vect{s}_t)$ cannot be computed in closed-form. We, therefore, resort to a variational approximation: $p_{\theta}(\vect{z}_t| \vect{s}_t) \approx q_{\psi}(\vect{z}_t| \vect{s}_t) $, with $q_{\psi}$ defined similarly as in \eqref{eq:encoder}. Overall, we have:
\begin{equation}
    p_{\theta}(\vect{z}, \bs{w} | \vect{s})\approx q_{\psi}(\vect{z}, \bs{w})=p_{\theta}(\bs{w} | \vect{s}, \vect{z}) q_{\psi}(\vect{z}| \vect{s}),
\end{equation}
where, $\bs{w}=\{w_1,\ldots, w_T\}$. We target a lower-bound on the data log-likelihood to learn $\widetilde{\Phi}$:
\begin{equation}
    \log p_{\theta}(\vect{s}) \ge \mathbb{E}_{q_\psi(\vect{z}, \bs{w})} \left\{ \log\frac{p_\theta(\vect{s},\vect{z}, \bs{w})}{q_\psi(\vect{z}, \bs{w})}\right\}\triangleq\mathcal{L}(\widetilde{\Phi};\vect{s}),
\end{equation}
which is simplified as
\begin{multline}
\mathcal{L}(\widetilde{\Phi};\vect{s})=\mathbb{E}_{q_\psi(\vect{z}, \bs{w})} \left\{ \log{p_\theta(\vect{s}|\vect{z}, \bs{w})}\right\}-\\\mathcal{D}_{\textsc{kl}}(q_\phi(\vect{z}|\vect{s}) \|p(\vect{z}))-\mathbb{E}_{q_\phi(\vect{z}|\vect{s})} \left\{ \mathcal{D}_{\textsc{kl}}(p_{\theta}(\bs{w}|\vect{z},\vect{s}) \|p(\bs{w}))\right\}.
\end{multline}
The first and third terms can be further simplified. This will bring us to the following final form:\footnote{Due to the limited space, we provide the detailed derivations in Supplementary Material available online: {\url{https://msaadeghii.github.io/files/stvae.pdf}}.}
\begin{multline}\label{eq:learnst}
        \mathcal{L}(\widetilde{\Phi};\vect{s}) = \sum_{t=1}^T \mathbb{E}_{q_\phi(\vect{z}_t|\vect{s}_t)}\{-\sum_{f=1}^F\log|{\sigma}_{\theta, f}^2(\vect{z}_t)| - \\(\alpha + F)\log(\beta + \sum_{f=1}^F \frac{|s_{ft}|^2}{{\sigma}_{\theta, f}^2(\vect{z}_t)})\} + \sum_{\ell=0}^{F-1}\log(\alpha+\ell) + \\ + \alpha\log\beta-\mathcal{D}_{\textsc{kl}}(q_\phi(\vect{z}|\vect{s}) \|p(\vect{z})).
\end{multline}

As in VAEs, we approximate the above expectation using a single sample $\vect{z}_t\sim q_\phi(\vect{z}_t|\vect{s}_t)$, followed by the reparametrization trick. The obtained objective function is then optimized over $\widetilde{\Phi}$ using a stochastic gradient-based optimizer. 
\subsection{Speech Enhancement}
\label{sec:se}
The observed noisy speech \ac{STFT} time frames are modeled as $ \vect{x}_t=\vect{s}_t+\vect{b}_t $, $t=1,\ldots,\tilde{T}$, where $ \vect{b}_t $ corresponds to background noise. For $\vect{s}_t$, the pre-trained generative model in \eqref{eq:propgenmodel} is used. For $ \bb_t $, the following \ac{NMF} based model is considered:
\begin{equation}
    \bb_t\sim \mathcal{N}_c(\boldsymbol{0}, \text{diag}(\Wb\bs{h}_t)),
    \label{eq:nmf}
\end{equation}
where, $ \Wb\in\Rbb_+^{F\times K} $, and $ \bs{h}_t $ is the $ t $-th column of $ \Hb\in\Rbb_+^{K\times \tilde{T}} $.

\subsubsection{Parameter estimation} To infer the model's parameters, i.e., $\phi = \lk \Wb, \Hb \rk$, we follow an EM approach, where in the expectation (E) step, the intractable posterior distribution $p(\vect{z}_t, w_{t} | \vect{x}_t)$ needs to be computed. As an approximation, we find only the modes, i.e., the points that maximize this distribution \cite{kameoka2019supervised,leglaive2020recurrent}:
\begin{equation}
    \vect{z}_t^*, w_t^* = \argmax_{\vect{z}_t, w_{t}}~\log p_\phi(\vect{z}_t, w_{t} | \vect{x}_t),
\end{equation}
or, equivalently,
\begin{equation}\label{eq:wz-up}
    \vect{z}_t^*, w_t^* = \argmax_{\vect{z}_t, w_t} ~\log p_\phi(\vect{x}_t|\vect{z}_t, w_t)+ \log p(\vect{z}_t) + \log p(w_t).
\end{equation}
It is straightforward to show that:
\begin{equation}\label{eq:px}
    p_\phi(\vect{x}_t|\vect{z}_t, w_t) = \mathcal{N}_c(\boldsymbol{0}, \diag(w_t^{-1}\bs{\sigma}_{\theta}^2(\vect{z}_t) + \Wb\bs{h}_t) ).
\end{equation}
Problem \eqref{eq:wz-up} is then solved via a first-order optimizer, e.g., Adam \cite{KingmaB15}. In the maximization (M) step, the parameters are updated by solving the following problem:
\begin{align}
     &\max_{\Wb, \Hb}\sum_{t}~ \E_{p_\phi(\vect{z}_t, w_{t} | \vect{x}_t)} \lk \log p_\phi(\vect{x}_t, \vect{z}_t, w_t)\rk \\
     \equiv&\max_{\Wb, \Hb}~ \sum_{t} \E_{p_\phi(\vect{z}_t, w_{t} | \vect{x}_t)} \lk \log p_\phi(\vect{x}_t| \vect{z}_t, w_t)\rk.
\end{align}
We approximate the above expectation using $\vect{z}_t^*, w_t^*$ as follows:
\begin{equation}
    \max_{\Wb, \Hb}~ \sum_{t} \log p_\phi(\vect{x}_t| \vect{z}_t^*, w_t^*).
\end{equation}
Substituting \eqref{eq:px} and pursuing the approach proposed in \cite{leglaive2018variance}, we obtain the following multiplicative update rules:
\begin{equation}
\mathbf{H} \leftarrow \mathbf{H} \odot \left( \frac{\mathbf{W}^\top \left( | \mathbf{X} |^{\odot 2} \odot \mathbf{V}^{\odot-2} \right)}{\mathbf{W}^\top \mathbf{V}^{\odot-1} } 
\right)^{\odot 1/2},
\label{updateH}
\end{equation}
\begin{equation}
\mathbf{W} \leftarrow \mathbf{W} \odot \left( \frac{\left( | \mathbf{X} |^{\odot 2} \odot \mathbf{V}^{\odot-2} \right) \mathbf{H}^\top}{\mathbf{V}^{\odot-1} 
\mathbf{H}^\top } \right)^{\odot 1/2},
\label{updateW}
\end{equation}
where $\odot$ denotes element-wise operation, $\mathbf{V} \in \mathbb{R}_+^{F \times \tilde{T}}$ is a matrix with columns $\vect{v}_{t} = (w_t^*)^{-1}\bs{\sigma}_{\theta}^2(\vect{z}_t^*) + \Wb\bs{h}_t$, and $\mathbf{X} \in \mathbb{C}^{F \times \tilde{T}}$ is a matrix with columns $ \vect{x}_t$. The overall inference algorithm iterates between \eqref{eq:wz-up}, \eqref{updateH}, and \eqref{updateW}.

\subsubsection{Speech estimation} Having learned $\phi^*=\lk\Wb^*,\Hb^*\rk$, the speech signal is estimated as the posterior mean $\hat{\vect{s}}_t = \E_{p_{\phi^*}(\vect{s}_t|\vect{x}_t)}\lk \vect{s}_t\rk, \forall t$, which can be equivalently written as
\begin{align}
    \hat{\vect{s}}_t &=\E_{p_{\phi^*}(\vect{z}_t^*, w_t^*|\vect{x}_t)}\lk \E_{p_{\phi^*}(\vect{s}_t|\vect{x}_t, \vect{z}_t^*, w_t^*)}\lk \vect{s}_t\rk\rk\nonumber\\
    &\approx\frac{(w_t^*)^{-1}\bs{\sigma}_{\theta}^2(\vect{z}_t^*)}{(w_t^*)^{-1}\bs{\sigma}_{\theta}^2(\vect{z}_t^*) + \Wb^*\bs{h}_t^*}\odot \vect{x}_t,
\end{align}
with element-wise division. 
\section{Related work}\label{sec:related}
The closest work to ours is \cite{takahashi2018student}, which presents a VAE for robust density estimation applications with a Gamma prior distribution on the variance of the Gaussian decoder. The parameters of this distribution are then modeled as functions of the latent codes, i.e., $\alpha(\vect{z})$ and $\beta(\vect{z})$, implemented by some DNNs. However, our approach is different, as we consider a variance model, $\bs{\sigma}_{\theta}^2(.)$, that is shared among all the data, and we instead consider separate scalar weights for each data point. We also do not model the Gamma parameters as functions of $\vect{z}$, because, it would highly complicate the optimization of $\vect{z}$ in the speech enhancement phase, i.e., \eqref{eq:wz-up}. Furthermore, in contrast to \cite{takahashi2018student}, we do not marginalize the weights and instead follow a variational approach, which is much more efficient.

\begin{table*}[t!]
\centering
	\caption{Average values of the input and output SI-SDR, PESQ, and STOI metrics for speech enhancement. The results are presented separately for the VAE models trained on clean data and outlier-contaminated data.}
 \renewcommand{\arraystretch}{1.2}
\resizebox{\textwidth}{!}{
\begin{tabular}{|l|c|c|c|c|c||c|c|c|c|c||c|c|c|c|c|}
\hline
 Metric & \multicolumn{5}{c||}{SI-SDR (dB)} & \multicolumn{5}{c||}{PESQ} & \multicolumn{5}{c|}{STOI} \\
\hline
{Noise SNR (dB)} & {-10} & {-5} & {0} & {5} & {10} & {-10} & {-5} & {0} & {5} & {10} & {-10} & {-5} & {0} & {5} & {10} \\ \hline\hline
Input (unprocessed) & -18.08 & -12.80 & -7.72 & -2.91  & 2.04    & 1.40 & 1.51 & 1.76 & 2.05 & 2.37 & 0.12        & 0.20       & 0.30       & 0.43       & 0.56                    \\ \hline
\multicolumn{16}{|c|}{Models trained on \textit{clean data}}\\
\hline
VAE-SE & -9.56 & -4.25 & 0.57 & 5.23 & 10.13 & {1.58} & 1.80 & 2.07 & 2.36 & 2.67 & 0.15         & 0.24        & 0.36      & 0.50       & 0.64 \\ \hline
StVAE-SE & \textbf{-8.92} & \textbf{-3.56} & \textbf{1.16} & \textbf{5.97} & \textbf{10.97} &  \textbf{1.61} & \textbf{1.85} & \textbf{2.17} & \textbf{2.47} & \textbf{2.73} & {{0.15}} & {\textbf{0.25}} & {\textbf{0.37}} & {\textbf{0.51}} & {\textbf{0.65}}\\ \hline
\multicolumn{16}{|c|}{Models trained on \textit{outlier-contaminated data}}\\
\hline
VAE-SE & -10.83 & -4.84 & 0.22 & 4.87 & 9.89  & 1.57 & 1.75 & 2.03 & 2.29 & 2.61 & 0.14                    & 0.23                  & 0.35                     & 0.49                     & 0.63 \\ \hline
StVAE-SE & \textbf{-9.23} & \textbf{-3.74} & \textbf{0.87} & \textbf{5.89} & \textbf{10.83} & \textbf{1.59} & \textbf{1.81} & \textbf{2.11} & \textbf{2.42} & \textbf{2.70} & {\textbf{0.15}} & {\textbf{0.24}}  & {\textbf{0.36}} & {\textbf{0.51}}  & {\textbf{0.65}} \\ \hline
\end{tabular}}
\label{tab:se_results}
\end{table*}
\section{Experiments}
\label{sec:exp}
\subsection{Setup}
In this section, we compare the performance of our proposed StVAE framework \eqref{eq:propgenmodel} against the standard VAE method based on \eqref{eq:genmodel} for both speech spectrogram auto-encoding and speech enhancement. The former consists of auto-encoding the \textit{clean} speech spectrogram using the trained VAE model to measure how well the input spectrogram is reconstructed, as also considered in \cite{bie2022unsupervised}. The reconstruction quality is measured based on the signal-to-noise ratio (SNR) in dB. Moreover, we plug the original phase into the reconstructed spectrogram and obtain the time-domain speech signal using inverse STFT. This is to evaluate the intelligibility and perceptual quality of the reconstructed speech signal in terms of the short-term objective intelligibility (STOI) measure~\cite{Taal2011stoi}, ranging in $[0,1]$, and the perceptual evaluation of speech quality (PESQ) score~\cite{Rix2001pesq}, ranging in $[-0.5,4.5]$, respectively. 

For speech enhancement performance evaluation, in addition to PESQ and STOI, we report the scale-invariant signal-to-distortion ratio (SI-SDR) \cite{le2019sdr} values. To have a fair comparison, the standard VAE-based speech enhancement (VAE-SE) considered as the baseline follows the same steps as those of StVAE-SE detailed in Section~\eqref{sec:se}. It should be mentioned that we did not include the VAE model proposed in \cite{takahashi2018student} as a baseline, because we could not get satisfactory results for the reasons mentioned in Section~\ref{sec:related}.

\subsection{Datasets}
For training the StVAE and VAE models, we used the speech data in the TCD-TIMIT corpus \cite{harte2015tcd}. This dataset contains speech utterances from 56 English speakers (39 for training, 8 for validation, and 9 for testing) with an Irish accent, uttering 98 different sentences, each with an approximate length of 5 seconds, and sampled at 16 kHz ($\sim$ 8 hours of data). The STFT of the speech data was computed with a 64 ms-long (1024 samples) sine window, 75$\%$ overlap, without zero-padding, which results in $F=513$.
\begin{table}[h!]
\centering
	\caption{Average values of the SNR, PESQ, and STOI metrics for speech spectrogram auto-encoding. }
 \renewcommand{\arraystretch}{1.2}
\begin{tabular}{|l|c|c|c|}
\hline
{Metric} & SNR (dB) & PESQ & STOI  \\ \hline
\multicolumn{4}{|c|}{Models trained on \textit{clean data}}\\
\hline
VAE & 6.94 & 3.29 & 0.85  \\ \hline
StVAE  &  \textbf{7.98} & \textbf{3.51} & \textbf{0.88} \\ \hline
\multicolumn{4}{|c|}{Models trained on \textit{outlier-contaminated data}}\\
\hline
VAE & {5.93} & 3.12 & 0.83  \\ \hline
StVAE & \textbf{7.15} & \textbf{3.32} & \textbf{0.86} \\ \hline
\end{tabular}
\label{tab:gen_results}
\end{table}

To test the speech enhancement performance, we used some noisy versions of the TCD-TIMIT dataset \cite{abdelaziz2017ntcd}, including six types of noise, namely \textit{\ac{LR}}, \textit{White}, \textit{Cafe}, \textit{Car}, \textit{Babble}, and \textit{Street}. For each noise type, we considered five noise levels: $-10$~dB, $-5$~dB, $0$~dB, $5$~dB, and $10$~dB. From each test speaker, we randomly selected 5 utterances for each  noise level and noise type, giving 1350 test samples.

Furthermore, to see how the two competing algorithms behave when the training data include some noise signals in addition to the clean speech data, we extended the training data by taking some noise data from the DEMAND dataset \cite{thiemann2013demand}, including \textit{STRAFFIC}, \textit{DWASHING}, \textit{SPSQUARE}, \textit{NRIVER}, \textit{TBUS}, \textit{NPARK}, and \textit{DKITCHEN}. The total amount of noise data is around $20 \%$ of the clean speech training data. As the speech generative model is supposed to be learned on only \textit{clean} data, including noise signals in the training set aims to measure the robustness of the learned models. We emphasize that, here, noise signals are \textit{not} added with clean speech signals to form mixtures. Instead, they are just intended to serve as some outlier training data.

\subsection{Model architecture} \label{sec:exp_methods}
The architectures of both StVAE and VAE follow the one proposed in \cite{leglaive2018variance}, consisting of an encoder and decoder each having a single fully-connected hidden layer with 128 nodes and hyperbolic tangent activation functions. The dimension of the latent space was set as $L=32$. 
\subsection{Parameters setting}
Both VAE models are trained with stochastic gradient descent (batch size of 128) using Adam. The learning rate is equal to $0.0001$. We used early stopping on the validation set with a patience of 20 epochs. The number of EM iterations for speech enhancement is set to 100. The learning rate for optimizing \eqref{eq:wz-up} is set to $0.005$, with 10 iterations.

Although $\alpha$ and $\beta$ in StVAE could be learned according to \eqref{eq:learnst}, we observed in our experiments that fixing these values during the whole training process leads to more stable and improved results. As such, we empirically set $\alpha = \beta = 100$, resulting in the mean and variance for the prior distribution of the weights equal to 1 and 0.01, respectively.

\subsection{Speech enhancement results}
The input (evaluated on unprocessed, noisy speech signals) and output (evaluated on the estimated speech signals) values of the speech enhancement metrics for different noise SNRs are reported in Table~\ref{tab:se_results}. Concerning the generative models learned on clean training data (without outlier data, i.e., noise signals), one can see that StVAE-SE outperforms VAE-SE in almost all the cases, demonstrating the efficiency of the proposed weighted variance Gaussian distribution compared to the standard, unweighted distribution. This is more noticeable for higher SNR levels. With respect to the outlier-contaminated data involving noise signals, we can also clearly see the advantage of StVAE. In addition, we note that StVAE-SE trained on outlier-contaminated data outperforms VAE-SE trained on clean data.

\subsection{Spectrogram auto-encoding results}
Table~\ref{tab:gen_results} summarizes the speech spectrogram auto-encoding results as a measure of reconstruction quality of the models. It can be seen that the proposed StVAE model performs considerably better than the standard VAE model, in terms of both reconstruction SNR and speech quality measures, PESQ and STOI. As also observed in the previous section, the results of StVAE when trained on outlier-contaminated data are better even than those of the VAE model trained on clean data. This confirms that the weighted variance model (Student's t-distribution) is a better fit for speech generative modeling.

It is important to mention that the performance improvements discussed above are achieved with very little additional computational overhead when compared to standard VAE-based speech generative modeling and enhancement. This comparison includes the generative models presented in equations \eqref{eq:genmodel} and \eqref{eq:propgenmodel}, as well as the new speech enhancement framework presented in equation \eqref{eq:wz-up}, which involves optimizing over the additional scalar-valued weight parameters.

\section{Conclusions}\label{sec:conc}
We presented a weighted variance Gaussian generative model for speech signals based on variational autoencoders. The proposed probabilistic generative model assumes a separate stochastic weight for each spectrogram time-frame with a Gamma prior distribution, providing a more flexible and effective modeling framework compared to the standard, unweighted Gaussian variance model. We also presented efficient parameter inference and speech enhancement methodologies. Our experimental results showed the superiority of the proposed model, both in terms of spectrogram auto-encoding reconstruction quality and speech enhancement results.

As future works, we plan to extend the proposed weighted variance generative model and speech enhancement frameworks to the dynamical VAE models \cite{bie2022unsupervised}, and to consider a Markovian dependency for the weights, which are not straightforward. This will allow for more efficient incorporation of the time-dynamics of the spectrogram time-frames, and consequently an improved performance. Furthermore, this will enable us to fairly compare the performance of the developed dynamical, weighted variance speech enhancement system with the supervised (discriminative), DNN-based approaches.

\bibliographystyle{IEEEtran}
\bibliography{mybib}

\end{document}